\documentclass{article}

\usepackage{arxiv}

\usepackage[utf8]{inputenc} 
\usepackage[T1]{fontenc}    
\usepackage{hyperref}       
\usepackage{url}            
\usepackage{booktabs}       
\usepackage{amsfonts}       
\usepackage{nicefrac}       
\usepackage{microtype}      
\usepackage{lipsum}
\usepackage{graphicx}
\usepackage{cite}
\usepackage{amsmath,amssymb,amsfonts}
\usepackage{algorithmic}
\usepackage{graphicx}
\usepackage{textcomp}
\usepackage{xcolor}
\usepackage[most]{tcolorbox}
\usepackage{lmodern}
\usepackage{enumitem}
\usepackage{url}
\usepackage[table]{xcolor}
\usepackage{longtable}
\def\BibTeX{{\rm B\kern-.05em{\sc i\kern-.025em b}\kern-.08em
    T\kern-.1667em\lower.7ex\hbox{E}\kern-.125emX}}

\usepackage{tcolorbox}
\usepackage{caption}
\usepackage{newfloat}

\DeclareFloatingEnvironment[
    fileext=lop,
    name=Prompt,
    listname=List of Prompts
]{prompt}

\title{From Text to DSL: Evaluating Grammar-Based Model Generation Using Open LLMs\thanks{in SOMET 2024:
23rd International Conference on Intelligent Software Methodologies,
2025} \thanks{This version includes corrections to several malformed references and incorrect arXiv links that appeared in the published conference version.}}

\author{
Junaid Baber\thanks{Corresponding author: junaid.baber@univ-grenoble-alpes.fr}
\and Nicolas Hili
\and Didier Schwab
\and Léo Challier
\and Cécilia Satrin
\\[1ex]
Univ. Grenoble Alpes, CNRS, Grenoble INP, LIG, 38000 Grenoble, France
}

\begin{document}
\maketitle
\begin{abstract}
Large Language Models (LLMs) have shown increasing potential in automating model-driven software engineering tasks, particularly in generating models conforming to Domain Specific Languages (DSLs) from natural language. While most existing approaches rely on large proprietary models, their high cost and limited deployability hinder broader adoption.

In this paper, we evaluate whether open-source LLMs of varying sizes (0.5B to 32B parameters) can generate DSL-conformant models using only few-shot prompting, without any fine-tuning. Our evaluation focuses on key model-driven engineering (MDE) requirements, including syntactic validity, semantic completeness, and inter-model reference consistency. 

We extend our prior work by moving from generating user interface models (referred to as "UI models" in this paper) over fixed, predefined data schemas ("data models") to generating both the UI and data models entirely from scratch. This shift serves two purposes: first, it highlights the LLM's ability to infer domain-specific relationships and maintain consistency across multiple interconnected models; second, it allows us to generalize earlier findings by testing DSL generation across models of different natures and structural roles. Our structured evaluation combines automatic parsing and expert feedback across 39 LLMs, revealing that several compact models (e.g., \texttt{gemma3:12b}, \texttt{mistral:7b-instruct}) approach or match the quality of much larger models.

These findings demonstrate the feasibility of using smaller, open-source LLMs for grammar-conformant DSL generation in MDE workflows, offering a cost-effective and deployable alternative to closed LLMs.
\end{abstract}


\section{Introduction}
Large Language Models (LLMs) have demonstrated significant potential in automating software engineering tasks, including code synthesis and model generation from natural language~\cite{white2023promptpatterns},~\cite{ busch2023chatgpt},~\cite{touvron2023llama}. In particular, the use of LLMs for generating models that conform to Domain-Specific Languages (DSLs) offers a promising path for increasing automation and accessibility in software development. DSLs, by design, are characterized by compact, domain-tailored grammars with strict syntactic and semantic rules. While this makes them ideal for automation, it also presents challenges for generative models, such as maintaining grammar conformance, resolving references across models, and ensuring completeness~\cite{hili2022lightweightlcdp}.

These challenges become even more pronounced in Model-Driven Engineering (MDE), where DSLs are used to define multiple interrelated models—such as domain models and interface models—that must conform to distinct metamodels and reference each other accurately. Ensuring consistency across these outputs is critical but difficult, as LLMs are not inherently structure-aware. Moreover, because DSLs are often underrepresented in LLM training corpora, few-shot prompting remains the most viable approach for guiding the generation of syntactically and semantically correct models.

Recent advances in Large Language Models (LLMs) have introduced a new paradigm for software generation via natural language~\cite{white2023promptpatterns, hagel2024noco}. These models can interpret high-level specifications and generate DSL code, UI elements, or even full-stack applications~\cite{busch2023chatgpt}. Work by Hagel et al.~\cite{hagel2024noco} demonstrated that LLMs like GPT-4 can be effectively integrated into Low Code Development Platforms (LCDPs) to synthesize textual models from user prompts, reducing time-to-model and improving usability. However, their approach assumes two significant constraints: (1) reliance on high-parameter, commercial LLMs (e.g., OpenAI GPT), and (2) use of pre-defined, fixed data models in the generation pipeline.

In this paper, we evaluate the capabilities of LLMs across different parameter scales —~from billion-scale proprietary models to open-source, resource-efficient ones like Mistral~\cite{jiang2023mistral7b} and LLaMA~\cite{touvron2023llama}~— for the task of generating DSL-conformant models. Our investigation focuses on whether few-shot prompting, without additional fine-tuning or retraining, is sufficient for these models to produce outputs that are syntactically valid, semantically complete, and structurally consistent within MDE workflows.

Building on our earlier work~\cite{hili2022lightweightlcdp}, where LLMs were used to generate UI models that referenced a fixed, predefined data model, we extend the challenge to include the automatic generation of the data model itself so that the approach can be generalized to more than one DSL. This shift significantly raises the difficulty, as it requires the LLM to infer application-specific domain concepts, attributes, and relationships directly from user-provided natural language descriptions. This expanded scope allows us to more comprehensively evaluate LLM capabilities across concept extraction, grammar compliance, and inter-model reference resolution.
Our contributions are threefold:

\begin{itemize}
 \item We present a structured evaluation of open-source LLMs—ranging from lightweight 0.5B models to larger 32B models—assessing their ability to generate DSL-conformant outputs using few-shot prompting, without any fine-tuning.

    \item We build on our prior work by shifting the focus from generating UI models over fixed schemas to the more challenging task of generating both data and UI models from scratch. This allows us to evaluate whether LLMs can infer application-specific domain concepts and preserve structural references across models.

    \item We design and conduct a two-pronged evaluation pipeline, including automated syntax validation and a human-centered semantic study, to benchmark model performance across criteria such as concept identification, completeness, and advanced feature inclusion.
\end{itemize}

By focusing on DSL-based modeling and conformance constraints, this study contributes to understanding the capabilities and limitations of LLMs in model-driven software engineering workflows. Our findings support the growing feasibility of using compact, open-source models in building intelligent modeling assistants  for different application domains, such as low-code or no-code platforms as explored in our previous paper~\cite{hagel2024noco}.

We explore the following research questions:

\begin{itemize}
    \item \textbf{RQ1:} \textit{How capable are small and mid-sized LLMs in generating DSLs from natural language specifications compared to large-scale LLMs?} \\
    This question investigates whether LLMs with significantly fewer parameters can produce DSL code that is syntactically and semantically correct, and whether they remain usable for non-technical users.

    \item \textbf{RQ2:} \textit{Can prompt engineering alone enable small LLMs to produce valid DSL models without the need for fine-tuning?} \\
    Here, we assess the effectiveness of structured and enriched prompts in guiding smaller LLMs to generate conformant DSL based models for application generation, without requiring retraining or fine-tuning the LLM.

\end{itemize}

\section{Related Work}

The rapid advancement of LLMs has necessitated new evaluation paradigms while simultaneously creating opportunities for their application in domain-specific language (DSL) creation and enhancement. Our work builds upon three intersecting research directions: evolving methodologies for LLM evaluation, LLM-assisted DSL development, and the integration of LLMs with DSL-based modeling workflows.

Traditional evaluation metrics such as accuracy and Bilingual Evaluation Understudy (BLEU) scores have proven inadequate for assessing modern LLMs' multifaceted capabilities. Recent work has developed more sophisticated approaches to address this limitation. The "LLM-as-Judge" paradigm, pioneered by Zheng et al.~\cite{zheng2023judging}, demonstrates that LLMs can assess text quality comparably to human evaluators when provided with detailed scoring rubrics. 

However, Wang et al.~\cite{wang2023adversarial} showed that carefully crafted adversarial in‑context demonstrations can mislead LLMs during in‑context learning, substantially degrading performance and exposing vulnerabilities in current evaluation methods.

Dynamic evaluation methodologies have emerged as a promising solution to prevent benchmark overfitting. Zhu et al.'s DyVal framework~\cite{zhu2023dyval} introduced on-the-fly task generation to continuously challenge model capabilities. This approach was extended by Li et al.~\cite{DynamicBench} through DynamicBench, a benchmark for evaluating LLMs’ ability to retrieve up‑to‑the‑minute information and generate concise reports by combining web search with local document retrieval. Game-theoretic evaluation protocols represent another significant advancement, with Khan et al.'s ZeroSumEval~\cite{khan2025zerosumeval} demonstrating how competitive interactions between models can reveal subtle robustness limitations. Despite these advances, current evaluation methodologies remain predominantly focused on general NLP tasks, leaving domain-specific evaluation—particularly for DSL-related applications—as an open research challenge.

\begin{figure}
\begin{center}
	 \includegraphics[width=0.9\textwidth]{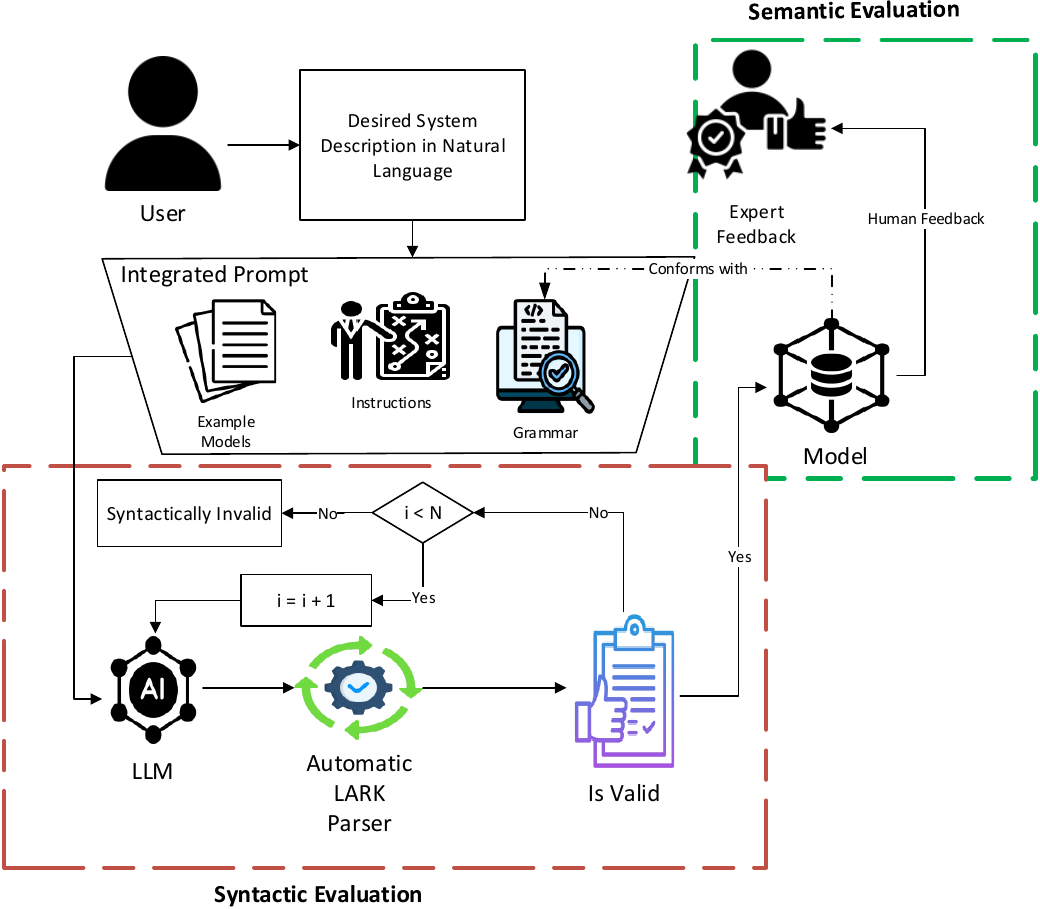}
\end{center}
\caption{Overview of the proposed model-based approach based on LLM for DSL creation.}
\label{fig:pipeline_overview}
\end{figure}

The application of LLMs to DSL creation has shown remarkable potential in democratizing domain-specific programming. Shraddha et al.~\cite{HYSYNTH} demonstrated that modern LLMs like GPT-4 can infer DSL grammars from structured input-output examples, though their accuracy significantly decreases for niche domains with limited training data. This limitation has spurred the development of hybrid approaches that combine neural generation with formal verification. DSLXpert~\cite{dslxpert2024} exemplifies this trend, proposing a prompt-based, grammar-conformant method for generating DSLs across domains without fine-tuning. Similarly, Kajal~\cite{torkamani2024kajal} explores how LLMs can reverse-engineer grammar rules from source code examples using prompt templates and few-shot learning.

Efforts in domain specialization also show promise. White et al.~\cite{white2023chemcoder} evaluated the chemistry knowledge of code‑generating LLMs on quantum chemistry problems and found that careful prompt engineering rather than additional domain specific pre‑training can significantly improve accuracy.  Parisotto et al.~\cite{parisotto2023neuro} proposed a neuro‑symbolic program synthesis approach that combines neural networks with symbolic search to synthesize programs in a small DSL from input–output examples.

Recent work on LLM-supported natural language-to-Bash translation~\cite{lin2024nl2bash} highlights how prompting techniques can enable reliable shell command generation. LangBiTe~\cite{lopez2024langbite} applies DSLs to specify and detect ethical biases in LLM outputs, while Pythoness ~\cite{levin2025pythoness} introduces an embedded DSL that guides LLMs to generate code from natural‑language or formal specifications via test‑driven development, improving the correctness of the generated code.

These examples collectively show the potential of LLM-DSL integration but also underscore current gaps: a focus on large-scale models, assumptions of fixed schemas, and limited attention to grammar conformance in few-shot settings. Our work attempts to address these issues by evaluating prompt-only DSL generation using small and mid-sized LLMs, supported by automatic syntax validation and dynamic model construction.

\section{Methodology and Experiments}

This section presents our experimental methodology for evaluating lightweight LLMs in the context of DSL generation. This approach is illustrated over the no-code platform case study as already explored in the previous paper. However, this approach can easily be generalized to other application domains. Our approach consists of a modular pipeline that includes generation of models of different natures, prompt engineering, LLM inference, and model evaluation.

\begin{table}
\centering
\caption{Evaluated LLMs grouped by architecture.  The gray-shaded models successfully generated syntactically valid DSLs recognized by the Lark parser. }
\begin{tabular}{|l|c|p{10cm}|}
\hline
\textbf{Model} & \textbf{Params} & \textbf{Description} \\
\hline
\multicolumn{3}{|c|}{\textbf{granitemoe}} \\
\hline
\rowcolor{gray!20} granite3-moe:3b~\cite{ibmgranite} & 3.4B & IBM's MoE-based LLM optimized for enterprise-grade reasoning and language tasks. \\
\rowcolor{gray!20} granite3-moe:latest~\cite{ibmgranite} & 1.3B & Lightweight version of Granite using mixture-of-experts for efficient inference. \\
\hline
\multicolumn{3}{|c|}{\textbf{olmo2}} \\
\hline
\rowcolor{gray!20} olmo2:13b~\cite{olmo} & 13.7B & Large-scale model by Allen AI for summarization, reasoning, and language generation. \\
olmo2:latest~\cite{olmo} & 7.3B & Balanced version of OLMo2 with improved inference efficiency. \\
\hline
\multicolumn{3}{|c|}{\textbf{llama}} \\
\hline
\rowcolor{gray!20} notus:latest & 7B & Instruction-tuned LLaMA 2 variant for RAG and general-purpose NLP. \\
\rowcolor{gray!20} dolphin-mistral:latest & 7B & Creative Mistral variant with uncensored output and conversational fine-tuning. \\
\rowcolor{gray!20} dolphin3:latest~\cite{cognitivecomputations2025dolphin3} & 8.0B & Chat-tuned model in the LLaMA family optimized for privacy and speed. \\
tinyllama:latest~\cite{zhang2024tinyllama} & 1B & Extremely compact LLaMA variant for edge and mobile applications. \\
\rowcolor{gray!20} codellama:13b~\cite{roziere2023code} & 13B & LLaMA-based model specialized for code generation and understanding. \\
\rowcolor{gray!20} codellama:latest~\cite{roziere2023code} & 7B & Smaller variant of CodeLlama for cost-effective code completion. \\
llama3.2:1b~\cite{touvron2023llama} & 1.2B & Tiny LLaMA 3.2 for on-device use and real-time applications. \\
llama3:latest~\cite{touvron2023llama} & 8.0B & General-purpose model from the LLaMA 3 series. \\
llama3.2:latest~\cite{touvron2023llama} & 3.2B & Mid-range LLaMA 3.2 variant with strong balance of speed and accuracy. \\
\rowcolor{gray!20} llama2:latest & 7B & Well-established LLaMA 2 model for open-source applications. \\
\rowcolor{gray!20} mistral:latest & 7.2B & Competitive dense decoder model with strong performance on reasoning benchmarks. \\
\rowcolor{gray!20} mistral:7b-instruct & 7.2B & Instruction-tuned version of Mistral for alignment tasks. \\
\rowcolor{gray!20} llama3.1:latest~\cite{touvron2023llama} & 8.0B & Enhanced LLaMA 3.1 for instruction following. \\
deepseek-r1:8b & 8.0B & LLaMA-based model tuned for logical reasoning and multi-task NLP. \\
\hline
\multicolumn{3}{|c|}{\textbf{stablelm}} \\
\hline
stable-code:latest~\cite{pinnaparaju2024stablecode} & 3B & Stability AI's compact model tailored for code completion and developer tools. \\
\hline
\multicolumn{3}{|c|}{\textbf{qwen2}} \\
\hline
\rowcolor{gray!20} openthinker:latest~\cite{openthoughts2025openthinker} & 7.6B & Qwen2-based model with strong reasoning and coding performance. \\
\rowcolor{gray!20} qwen2.5-coder:14b~\cite{hui2024qwen} & 14.8B & Code-specialized LLM with 128K context support. \\
qwen2.5-coder:0.5b~\cite{hui2024qwen} & 494.03M & Lightweight code generation model in the Qwen2.5 family. \\
\rowcolor{gray!20} qwen2.5-coder:1.5b~\cite{hui2024qwen} & 1.5B & Compact Qwen2.5 model for fast, local code synthesis. \\
\rowcolor{gray!20} qwen2.5-coder:latest~\cite{hui2024qwen} & 7.6B & Efficient and scalable Qwen2.5-Coder release. \\
\rowcolor{gray!20} qwen2.5-coder:3b~\cite{hui2024qwen} & 3.1B & Mid-size variant tuned for balanced code understanding. \\
qwq:latest~\cite{qwq2024} & 32.8B & Massive Qwen2-style LLM with unclear internal details. \\
\rowcolor{gray!20} deepseek-r1:14b~\cite{deepseek2025r1} & 14.8B & Large DeepSeek-R1 variant for high-complexity reasoning. \\
\rowcolor{gray!20} deepseek-r1:latest~\cite{deepseek2025r1} & 7.6B & General-purpose Qwen-style model for logical NLP. \\
\rowcolor{gray!20} qwen2.5:latest~\cite{hui2024qwen} & 7.6B & Base Qwen2.5 model with optimized inference and language capability. \\
deepseek-r1:1.5b~\cite{deepseek2025r1} & 1.8B & Small variant with strong reasoning accuracy and efficiency. \\
\hline
\multicolumn{3}{|c|}{\textbf{phi2/phi3}} \\
\hline
phi:latest & 3B & Transformer model trained on curated web and synthetic content. \\
\rowcolor{gray!20} phi3:14b~\cite{abdin2024phi3} & 14.0B & High-capacity Phi model tuned for math and logic tasks. \\
\rowcolor{gray!20} phi3:latest~\cite{abdin2024phi3} & 3.8B & Compact version of Phi-3 for efficient reasoning. \\
\rowcolor{gray!20} phi4:latest~\cite{abdin2025phi4} & 14.7B & State-of-the-art Phi model with superior instruction tuning. \\
\hline
\multicolumn{3}{|c|}{\textbf{gemma}} \\
\hline
gemma:2b~\cite{gemma2024open} & 3B & Open-weight Gemini-based model for general NLP tasks. \\
\rowcolor{gray!20} gemma:latest~\cite{gemma2024open} & 9B & Larger Gemma model optimized for robust understanding. \\
\hline
\multicolumn{3}{|c|}{\textbf{gemma3}} \\
\hline
\rowcolor{gray!20} gemma3:4b~\cite{gemma2025gemma3} & 4.3B & Mid-size Gemma3 for fast and expressive reasoning. \\
\rowcolor{gray!20} gemma3:12b~\cite{gemma2025gemma3} & 12.2B & High-capacity Gemma3 model tuned for comprehension tasks. \\
gemma3:1b~\cite{gemma2025gemma3} & 999.89M & Lightweight model suitable for edge devices. \\
\hline
\end{tabular}
\label{tab:llm_overview}
\end{table}

\subsection{Overview of the Evaluation Pipeline}

Figure~\ref{fig:pipeline_overview} illustrates the overall workflow. The evaluation process begins with a user-provided natural language specification describing the target application (e.g., a conference planner or an ice cream parlor website). This specification is processed in two main stages: first, the generation of a structured data model capturing the core entities, attributes, and relationships; and second, the construction of a prompt that incorporates this model to drive model synthesis via an LLM.

The  output DSL code is subsequently verified for syntactic correctness through a LARK parser and validated  for semantic alignment through human assessment. This will be detailed in Section 4.

\subsection{LLM Selection and Configuration}

We selected a diverse set of open-source LLMs spanning various families, architectures, and parameter sizes. These include models from the LLaMA, Qwen2, Phi, Gemma, and GraniteMoE families, among others (see Table~\ref{tab:llm_overview}). Our selection captures variations in scale (from 0.5B to 32B), training objectives (instruction-tuned vs. base), and model type (code-focused vs. general-purpose).

All models were used without fine-tuning. Where possible, inference was performed locally using quantized versions (e.g., GGUF), otherwise via public APIs.

\begin{prompt}
\caption{Prompt for generating a DSL model for an online ice cream parlor.}
    \begin{tcolorbox}[colback=gray!5!white, colframe=gray!75!black, title=Example Prompt]

\small

You are a \textbf{DSL generator}. Given a user's intent to create a website or system, you must generate a \textbf{DSL-style data model}.

\textbf{Rules:}
\begin{itemize}[leftmargin=1.5em]
  \item Only output the \textbf{DSL} — no explanations, comments, or extra text.
  \item Follow the \textbf{grammar} for generation. The output must comply with the rules given below.
  \item Analyze the user input and \textbf{expand the idea} to ensure the data model covers all necessary aspects, entities, and relationships.
  \item Use the exact \textbf{syntax and style} shown in the example:
  \begin{itemize}
    \item \texttt{main concept} for the central concept
    \item \texttt{concept} for entities
    \item \texttt{enum} for enumerations
    \item \texttt{one}, \texttt{some}, \texttt{lone} for cardinality
    \item \texttt{-->} for one-way references
    \item \texttt{<>-->} for many-to-many or one-to-many bidirectional associations
    \item \texttt{isId} to mark identifiers
    \item \texttt{subset of} to indicate constrained relationships
    \item Provide default values for enums or primitives where applicable
  \end{itemize}
\end{itemize}
\textbf{Grammar: [grammar]:} \\
\textbf{Example:} 

\hspace{2em}\textbf{[Example User Input]} 

\hspace{2em}\textbf{[Example DSL]} \\
\textbf{User Input:} Now, generate a DSL for: \texttt{"I want to create the website for online icecream parlor"}

\end{tcolorbox}
\end{prompt}

\subsection{Prompt and Data Model Integration}

The initial step involves converting a user's natural language specification into a structured data model. This model includes the key concepts that define the application domain, such as entities (e.g., \texttt{Event}, \texttt{User}, \texttt{IceCreamFlavor}), their attributes (e.g., \texttt{name}, \texttt{date}, \texttt{type}), and relationships (e.g., one-to-many, many-to-many). This transformation is carried out by prompting an LLM to infer and extract the conceptual structure implied by the user's input. The resulting model is manually reviewed to ensure it is syntactically valid and semantically consistent with the specification.

Once the data model is created, it becomes an integral part of the DSL generation prompt. A unified prompt template was developed to ensure consistent and fair comparison across all LLMs. This template includes the DSL grammar (written in Xtext-style syntax), one or more example DSL models for context, the generated data model, and the user's original specification, the example prompt is shown in Prompt~1. The prompt is framed to instruct the LLM to generate DSL code without commentary—strictly conforming to the syntax and constructs defined in the grammar. By standardizing the prompt content and format, we ensure that model comparisons are based on their generalization and reasoning capabilities rather than training artifacts or fine-tuning.

To further support grammar-conformant and semantically rich DSL generation, the prompt is designed to explicitly instruct the LLM on both syntax and generation behavior. It positions the model as a ``DSL generator,'' not a generic assistant, and defines strict output rules--such as avoiding natural language, adhering to cardinality and type syntax, and including identifiers and subset relationships. This mirrors the requirements of our DSL grammar and MDE practices more broadly.

The prompt includes an example, reminders of key syntax conventions (e.g., \texttt{main concept}, \texttt{<>-->}, \texttt{isId}), and a directive to semantically expand under-specified inputs. This structure follows prompt engineering best practices~\cite{white2023promptpatterns}, and supports both consistency and generalizability across models of varying sizes. By embedding these constraints directly in the prompt, we ensure that even models not fine-tuned on DSLs can generate usable outputs through structured prompting alone.

\begin{figure}
\begin{center}
\begin{tabular}{cc}
    \includegraphics[width=0.5\textwidth]{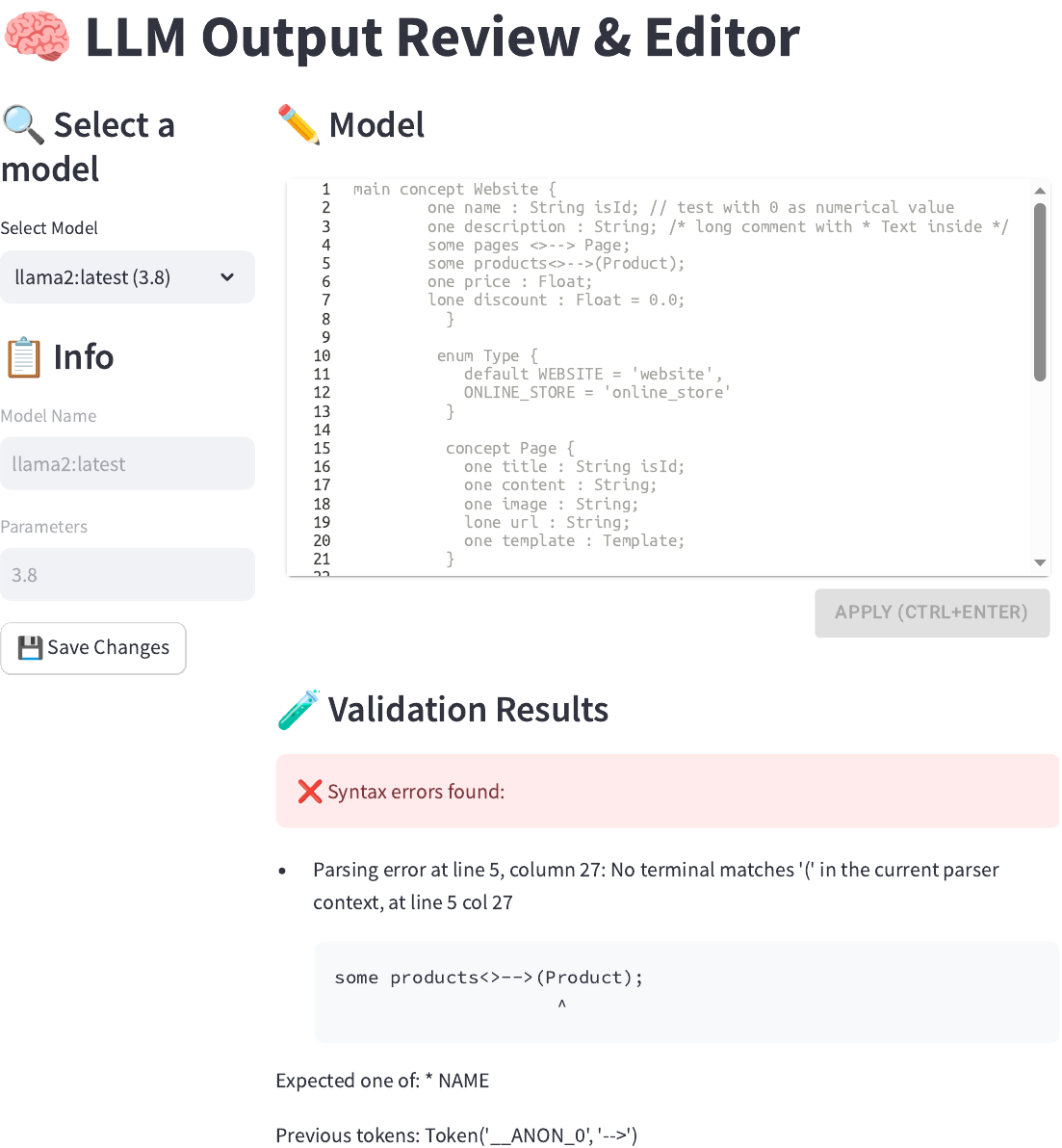} &    \includegraphics[width=0.5\textwidth]{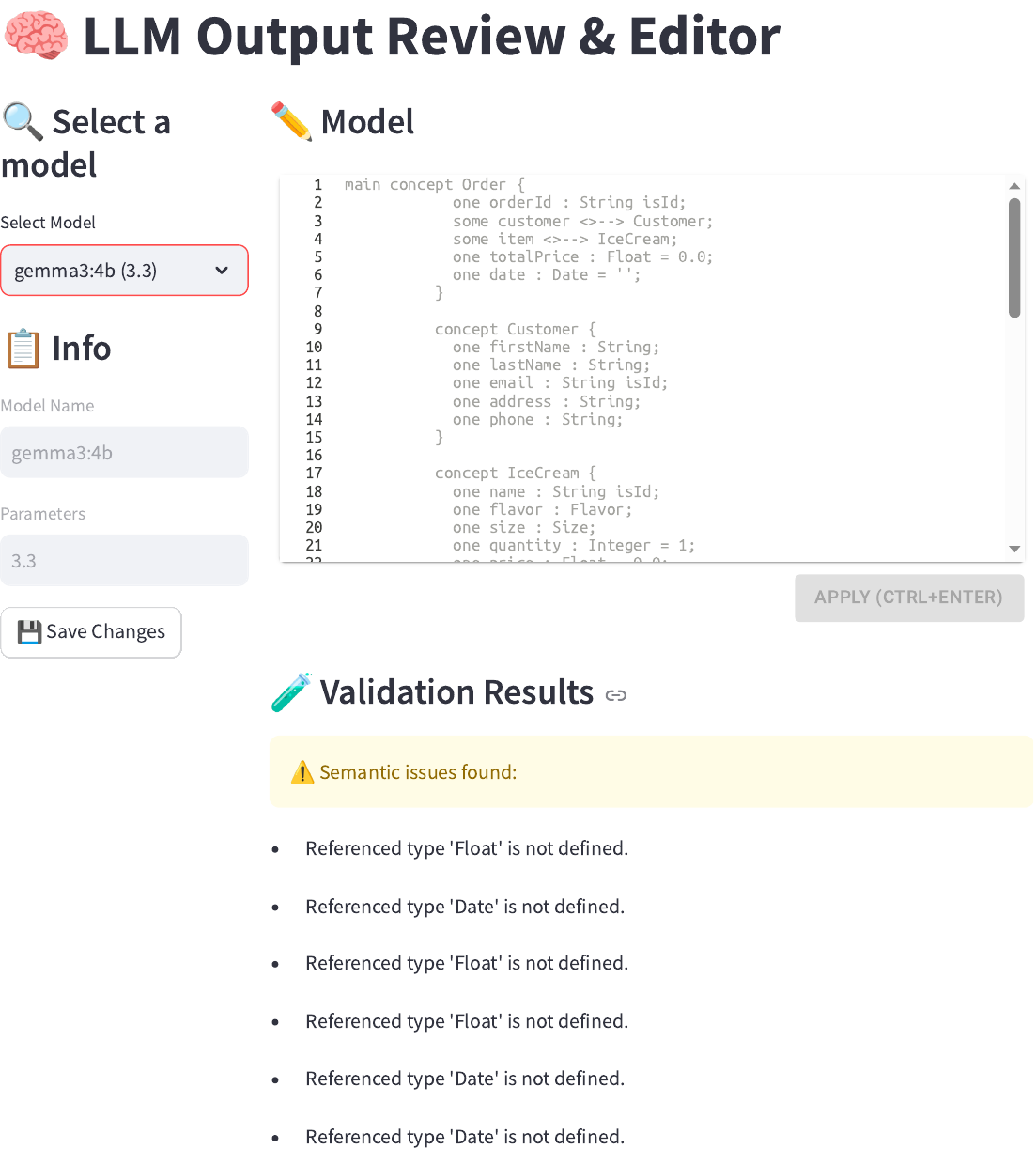}\\ (a) & (b) 

\end{tabular}
\end{center}
\caption{Preview of LARK based parser implementation that identifies syntax errors with token level precision and provides context-specific error reporting, (a) shows syntax errors, (b) shows semantic issues identified.}

\label{fig:feedback_lark}
\end{figure}

\subsection{Evaluation Setup}

We designed a test suite of natural language scenarios drawn from multiple domains, reflecting typical use cases in low-code and model-driven development platforms. These scenarios were processed through our pipeline using all selected LLMs to evaluate their ability to generate structured and conformant DSL models.

\subsubsection{Syntactic Validation}

To assess syntactic validity, we implemented a parser using the LARK framework. The parser is built upon a comprehensive DSL grammar and attempts to parse each generated output using a bottom-up LALR (Look-Ahead Left-to-Right) parser. The parser identifies syntax errors with token-level precision and provides context-specific error reporting, as illustrated in Figure~\ref{fig:feedback_lark}. In addition to grammar conformity, the parser incorporates a semantic validation layer that checks for unresolved type references, invalid inheritance targets, and ill-formed subset relationships. This goes beyond token correctness and ensures conformity with metamodel-style constructs typical in MDE workflows.

To improve reliability, we implemented an automatic verification loop that attempts to regenerate DSL output for each model up to \( N \) times in the event of a parsing failure. Many LLMs produce syntactic issues in initial attempts, such as missing semicolons or the use of unsupported symbols. When such errors are detected by the LARK parser, feedback is generated automatically and the prompt is re-issued to the model.

Given the non-deterministic behavior of LLMs, particularly when using higher temperature values, we leveraged a two-phase temperature setting strategy. In the first iteration, we used the default temperature setting (typically around 0.7--1.0) to allow the model to expand on the input and exercise creativity. If the output failed to parse, subsequent retries (up to \( N-1 \)) were made with a reduced temperature of 0.1, encouraging more deterministic and consistent responses. This reduction minimizes variation and increases the likelihood of producing syntactically valid code based on earlier feedback.

If all \( N \) attempts fail, the model is marked as syntactically invalid for that specific scenario. Models that could not generate any syntactically valid DSL after \( N \) attempts were excluded from the semantic evaluation phase, as their outputs could not be reliably interpreted or assessed. This approach ensures that the semantic evaluation focuses only on structurally viable outputs, in line with RQ1 and RQ2.

\subsubsection{Semantic Evaluation}

To complement automated validation, we conducted a human assessment study involving three domain experts with advanced knowledge of DSLs, modeling, and language semantics. Each expert was asked to evaluate the semantic quality of DSL models generated by different LLMs for the same natural language scenario.

The models were evaluated across four criteria: \textit{Semantic Correctness}, \textit{Concept Identification}, \textit{Completeness}, and \textit{Advanced Features}. The “Advanced Features” dimension assessed whether the generated models went beyond basic entity and relationship structures to include rich, domain-specific constructs—such as customer ratings (e.g., \texttt{Review}, \texttt{Rating}), promotions (e.g., \texttt{Discount}, \texttt{Promotion}), delivery tracking, or user personalization.

Scores were assigned using a 5-point Likert scale, where 1 indicated poor performance and 5 indicated excellent semantic coverage. This manual evaluation approach allowed us to capture aspects of DSL quality that cannot be fully assessed through structural or syntactic validation alone.

\begin{figure}
\begin{center}
\begin{tabular}{cc}
    \includegraphics[width=0.45\textwidth]{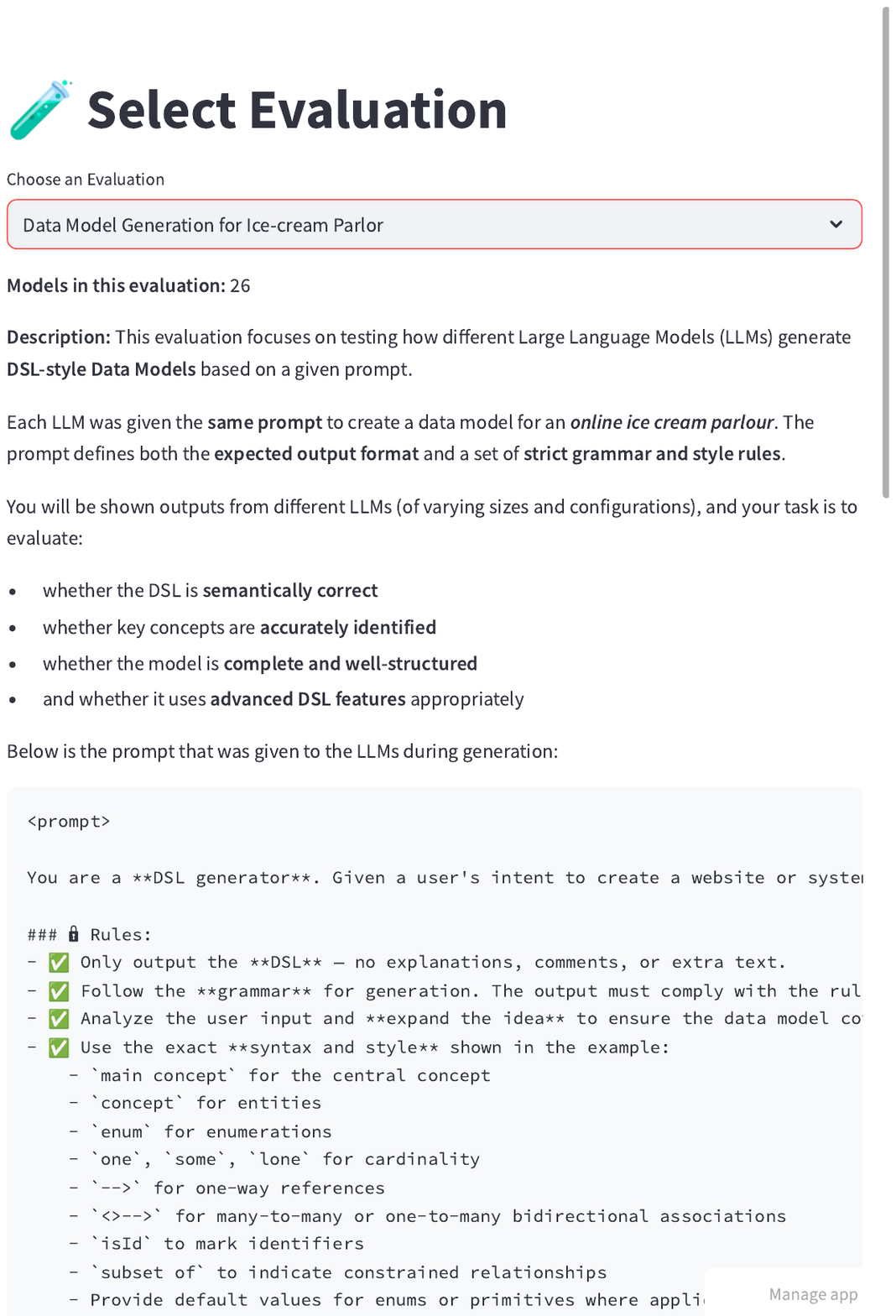} &    \includegraphics[width=0.45\textwidth]{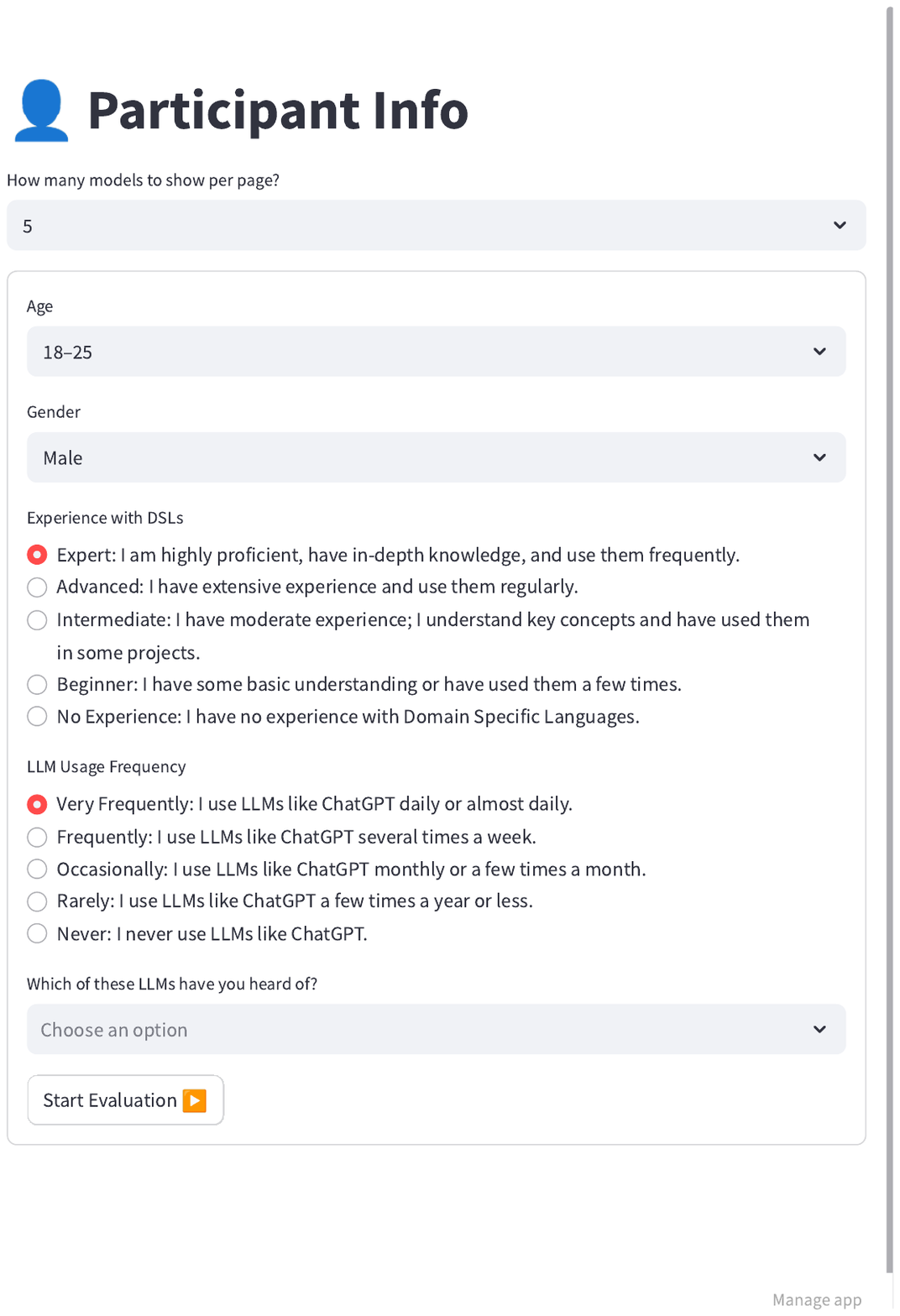}\\ (a) & (b) \\
    \multicolumn{2}{c}{\includegraphics[width=0.45\textwidth]{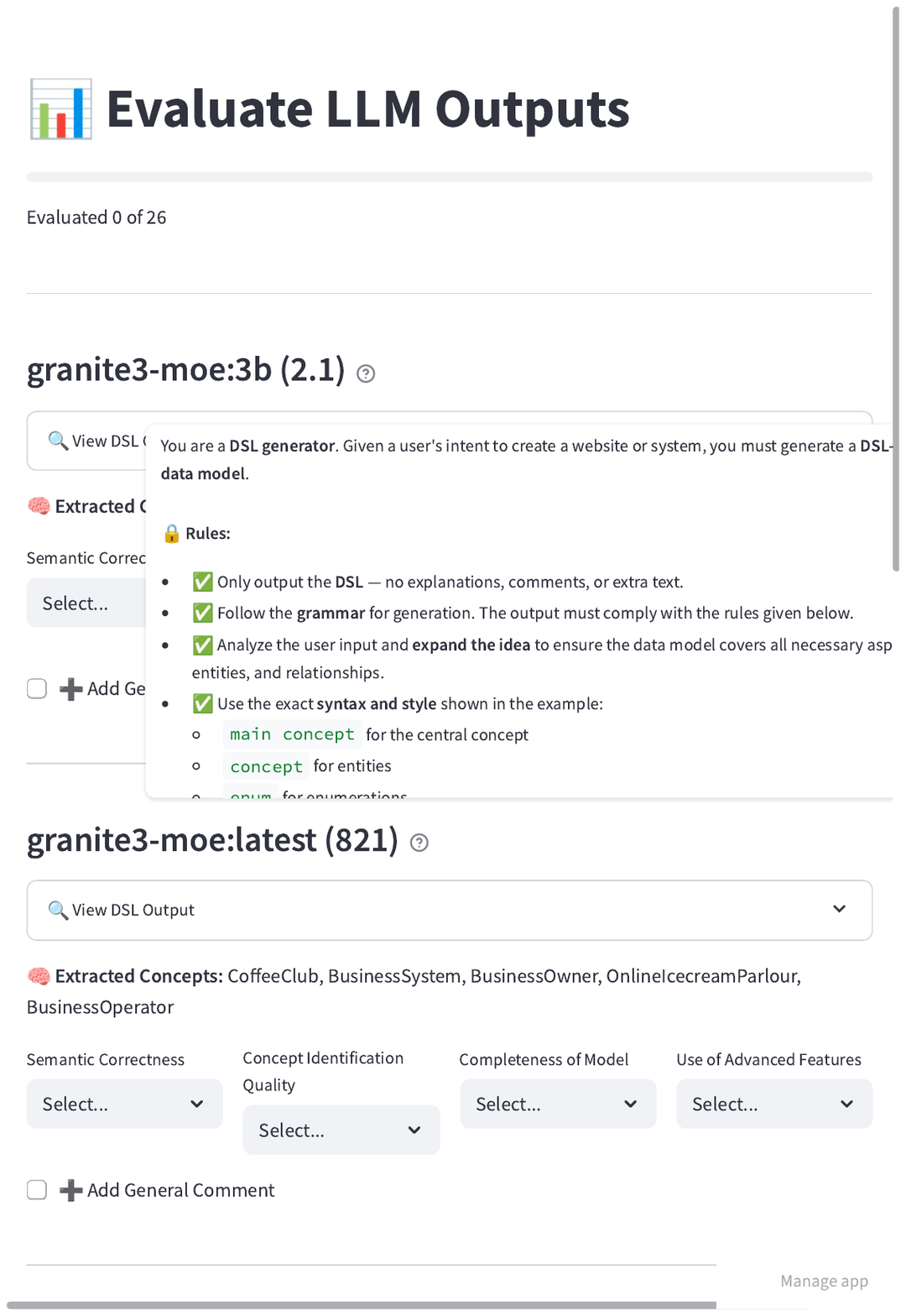}} \\  \multicolumn{2}{c}{(c)}

\end{tabular}
\end{center}
\caption{Overview of the human evaluation interface. Panel (a) displays the available evaluation experiments, panel (b) captures participant demographic information, and panel (c) presents the model outputs along with corresponding evaluation criteria. Users can access the original prompt used for model generation by clicking the help icon.}

\label{fig:feedback}
\end{figure}

To streamline and standardize the evaluation, we developed a web-based feedback platform~\footnote{\url{https://baberjunaid-llm-evaluation-feedback-feedback-app-u5vsif.streamlit.app/}}. Participants selected tasks, entered demographic details (age, gender, DSL experience, and LLM usage frequency), and then reviewed generated DSL outputs with syntax highlighting and extracted concept summaries. Each model was rated independently across the four criteria, and participants could optionally provide qualitative comments. Evaluation progress was tracked with a progress bar, and feedback was submitted to a centralized backend, as illustrated in Figure~\ref{fig:feedback}.

\begin{figure}
\begin{center}
\begin{tabular}{cc}
    \includegraphics[width=0.5\textwidth]{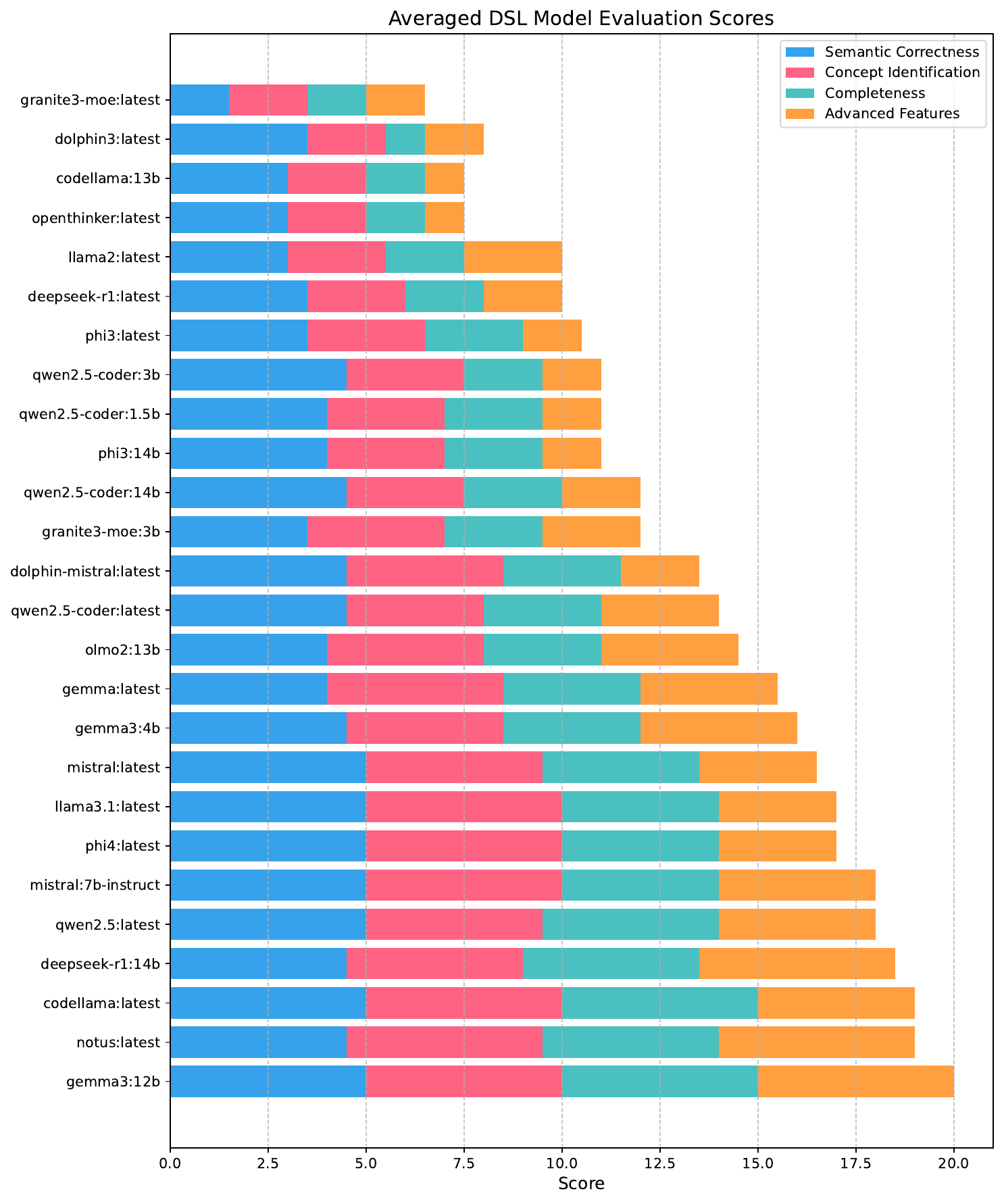} & \includegraphics[width=0.5\textwidth]{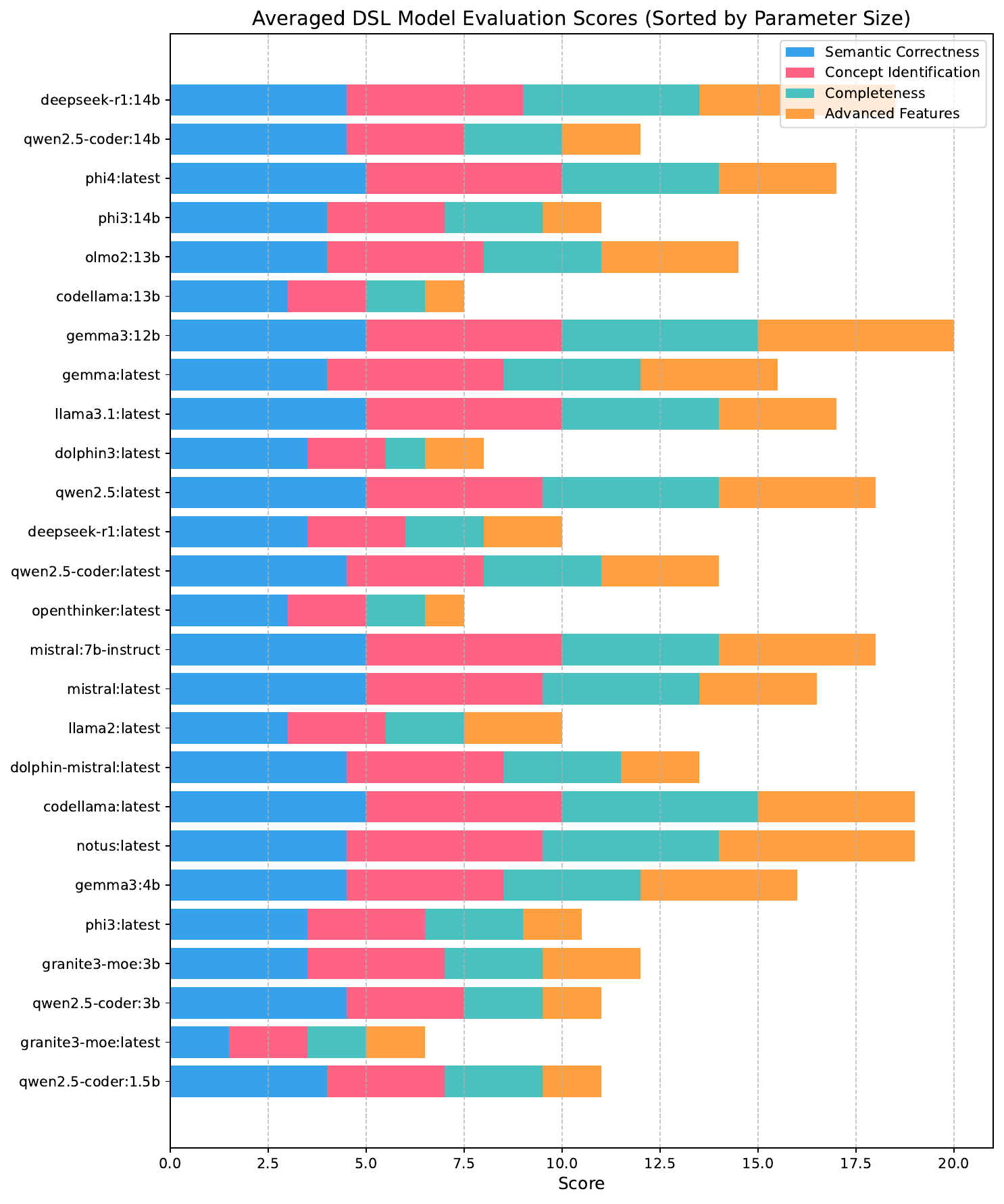}

\end{tabular}
\end{center}
\caption{Stacked bar chart of averaged evaluation scores for 26 DSL models designed for an online ice cream parlour, (a) shows models by score and (b) shows model by size.}
\label{fig:results1}
\end{figure}

\begin{figure}
\begin{center}
\begin{tabular}{c}
  
    \includegraphics[width=0.75\textwidth]{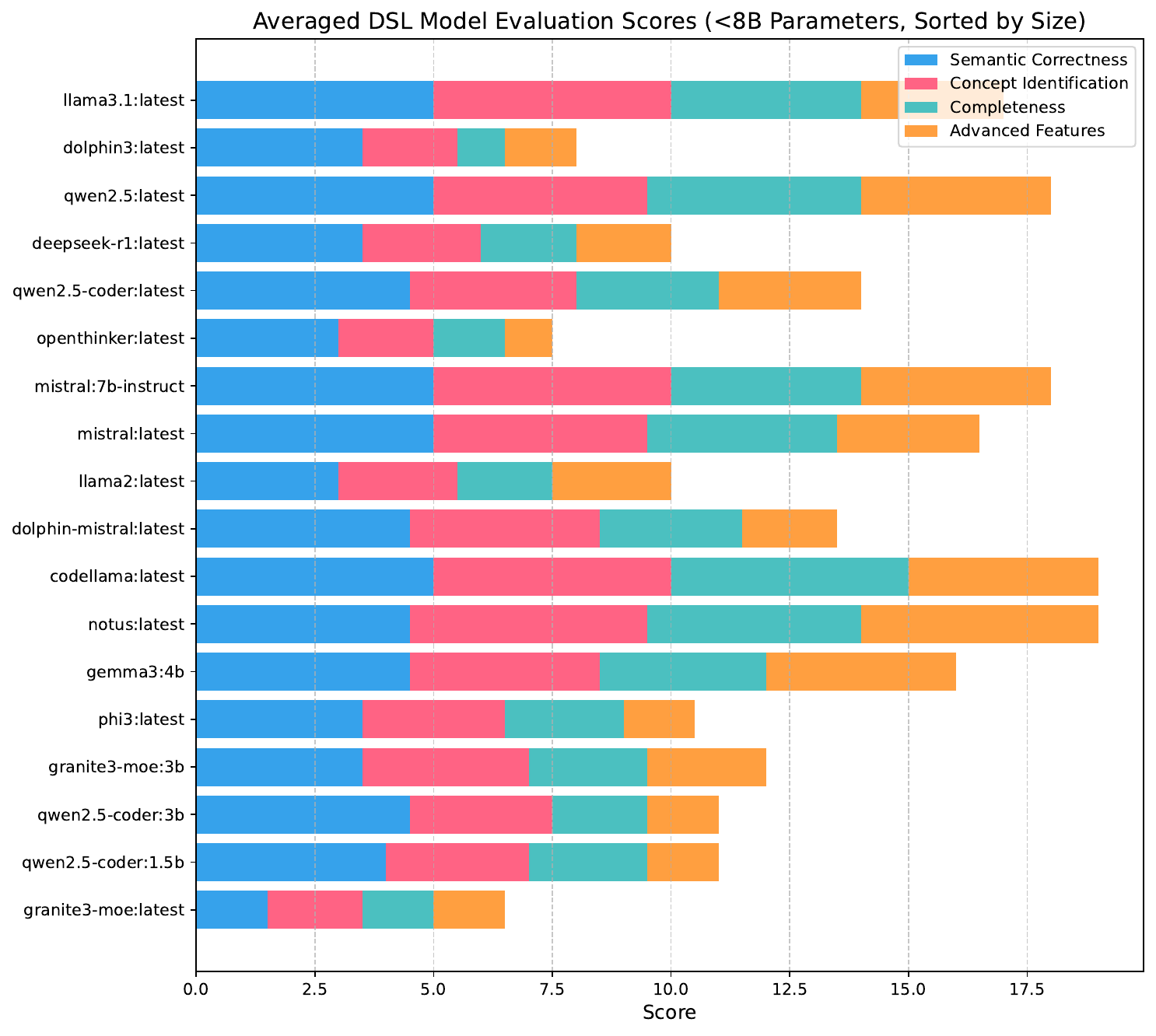}\\

\end{tabular}
\end{center}
\caption{Stacked bar chart of averaged evaluation scores for 18 DSL models with fewer than 8 billion parameters, designed for an online ice cream parlour.}
\label{fig:results2}
\end{figure}

\section{Results and Analysis}

We evaluated 39 LLMs to determine their ability to generate DSL models that are both syntactically valid and semantically meaningful. Only 26 of these models produced at least one syntactically correct output within the allowed retry limit, and these were subsequently included in the semantic evaluation.

To visualize the outcomes, we present stacked bar charts that reflect both syntactic validation and semantic performance. Figure~\ref{fig:results1}~(a) shows all 26 valid models, normalized to a maximum semantic score of 20, illustrating how performance is distributed across four evaluation criteria. Models such as \texttt{gemma3:12b} and \texttt{mistral:7b-instruct} performed well across all categories, while \texttt{granite3-moe:latest} demonstrated lower semantic coverage.

Figure~\ref{fig:results1}~(b) organizes these models by parameter size. Although larger models like \texttt{phi4:latest} scored highly, several smaller models, including \texttt{notus:latest} and \texttt{codellama:latest}, achieved comparable or better results. This suggests that larger parameter counts are not necessarily indicative of superior DSL generation performance.

Figure~\ref{fig:results2} focuses on models with parameter sizes \( \leq 8\text{B} \), highlighting that smaller models can still produce competitive outputs when guided with grammar-conformant prompts and retry-based validation.

Top performers included \texttt{gemma3:12b}, \texttt{notus:latest}, and \texttt{mistral:7b-instruct}, all of which demonstrated consistent quality across the evaluated dimensions. In contrast, models such as \texttt{granite3-moe:latest}, \texttt{dolphin3:latest}, and \texttt{codellama:13b} often lacked structural completeness and produced less detailed outputs.

Some compact models like \texttt{phi3:latest} and \texttt{qwen2.5-coder:1.5b} showed promising results, especially when retry mechanisms were applied at reduced temperature. Instruction-tuned models generally produced more consistent and accurate outputs. However, the inclusion of advanced semantic features varied significantly—some models incorporated enums, subset relations, and cross-entity references, while others generated only the most essential entity definitions.

These results demonstrate that prompt engineering and structured validation are effective strategies for enabling even smaller LLMs to generate usable DSL models. With appropriate prompt design and retry handling, low-resource models can support no-code development tasks that traditionally depend on larger, commercial systems.

\textit{RQ1} is addressed through our semantic evaluation results (Figures~\ref{fig:results1} and~\ref{fig:results2}), which show that several small and mid-sized models (e.g., \texttt{mistral:7b-instruct}, \texttt{notus:latest}) achieve DSL generation performance comparable to larger models. 
\textit{RQ2} is supported by the fact that all models were used without fine-tuning, relying only on prompt engineering. Our syntactic and semantic validation pipeline demonstrates that with carefully designed prompts, even small models can generate conformant and meaningful DSL outputs.
\section{Conclusion}

This paper explored the capabilities of large language models for generating domain-specific models that conform to strict DSL grammars, a task critical for enabling no-code platforms grounded in model-driven engineering. We evaluated 39 LLMs of varying sizes and architectures, focusing on their ability to generate syntactically valid and semantically complete DSL models using prompt engineering alone.

Our results show that many small and mid-sized open-source models are capable of producing valid and useful outputs when guided with carefully constructed prompts and supported by retry-based feedback loops. Instruction-tuned models consistently performed better, and semantic coverage improved when models were prompted to expand the user's input thoughtfully.

While high-parameter models performed reliably, our findings suggest that smaller models can be practical alternatives in contexts where privacy, cost, or deployment constraints limit the use of commercial LLMs. Future work may explore fine-tuning small LLMs to match or exceed the performance of mid-sized models, as well as investigate multi-turn refinement workflows and tighter integration with verification tools to improve correctness and reliability.

All source code and experimental setups are available at our GitHub repository.\footnote{\url{https://github.com/baberjunaid/LLM_DSL_Evaluation}}

\bibliographystyle{unsrt}  

\bibliography{mybib}

\end{document}